\documentclass[twocolumn,showpacs,preprintnumbers,amsmath,amssymb]{revtex4}
\usepackage{graphicx} 
\usepackage{bm}
\usepackage{dcolumn}
\usepackage{epstopdf}
\usepackage{natbib}

\begin{document}

\title{Density of states in the bilayer graphene with the excitonic pairing interaction}

\author{V. Apinyan\footnote{Corresponding author. Tel.:  +48 71 3954 284; E-mail address: v.apinyan@int.pan.wroc.pl.}, T. K. Kope\'{c}}
\affiliation{Institute for Low Temperature and Structure Research, Polish Academy of Sciences\\
PO. Box 1410, 50-950 Wroc\l{}aw 2, Poland \\}

\date{\today}

\begin{abstract}
%
In the present paper, we consider the excitonic effects on the single particle normal density of states (DOS) in the bilayer graphene (BLG). The local interlayer Coulomb interaction is considered between the particles on the non-equivalent sublattice sites in different layers of the BLG. We show the presence of the excitonic shift of the neutrality point, even for the noninteracting layers. Furthermore, for the interacting layers, a very large asymmetry in the DOS structure is shown between the particle and hole channels. At the large values of the interlayer hopping amplitude, a large number of DOS at the Dirac's point indicates the existence of the strong excitonic coherence effects between the layers in the BLG and the enhancement of the excitonic condensation. We have found different competing orders in the interacting BLG. Particularly, a phase transition from the hybridized excitonic insulator phase to the coherent condensate state is shown at the small values of the local interlayer Coulomb interaction.
\end{abstract}

  \maketitle

\renewcommand\thesection{\arabic{section}}

\section{\label{sec:Section_1} Introduction}
%
The problem of excitonic pair formation in the graphene and bilayer graphene (BLG) systems is one of the longstanding and controversial problems in the modern solid state physics \cite{cite_1,cite_2,cite_3,cite_4,cite_5,cite_6,cite_7,cite_8,cite_9,cite_10,cite_11,cite_12}. The chiral invariance of the free quasiparticle Hamiltonian, combined with electron Coulomb interaction term brought the idea about the possibility of spontaneous chiral symmetry breaking (CSB) in a single layer and bilayer graphene, reflecting in the form of the gapped states in the fermionic quasiparticle spectrum \cite{cite_1, cite_13, cite_14,cite_15,cite_16,cite_17,cite_18}. The excitonic gap equation has been derived in the Refs.\cite{cite_5, cite_6, cite_7, cite_8, cite_9, cite_10, cite_11} at the Bardeen-Cooper-Schrieffer (BCS) limit of the excitonic transition scenario. The general idea used there is based on the supposition of the weak electronic correlations in the BLG, due to its large number of the fermionic flavors. It has been suggested in Refs.\cite{cite_1, cite_16, cite_17} that even a partial account of the wave vector or energy dependence of the gap function rules out the constant gap solution in the BCS limit. Meanwhile, it has been shown that even undoped graphene can provide a variety of electron-hole type pairing CSB orders especially for the strong Coulomb coupling case \cite{cite_16, cite_17}, which renders the treatments in \cite{cite_1, cite_13, cite_14,cite_15, cite_18} to be obscure. 

Concerning the excitonic condensation, the proper inclusion of the chemical potential fluctuation effects on the excitonic pairing gap and condensation in the BLG system, and also the role of the average chemical potential on the excitonic effects have been discussed in Refs.\cite{cite_10, cite_11, cite_12}. A very large excitonic gap, found in a sufficiently broad interval of the repulsive interlayer Coulomb interaction parameter, given in Ref.\cite{cite_12}, suggests that the level of intralayer density and chemical potential fluctuations in the BLG system, discussed in Ref.\cite{cite_11}, are sufficiently small and do not affect the robust excitonic insulator state in the BLG. The excitonic condensation in the single BLG is principally possible in the case of the static interlayer screening regime \cite{cite_10}. 

Moreover, it has been shown that there exists a critical value of the interlayer hopping amplitude $\gamma_1$ \cite{cite_15}, which provides an interesting energy cutoff, below which the electron-hole correlations do not drive the system towards the CSB excitonic transition. More recently, it has been shown \cite{cite_19, cite_20} that the single-particle coherent density of states (DOS) in the usual two-dimensional (2D) \cite{cite_19} and three-dimensional (3D) \cite{cite_20} semiconducting systems at the zero temperature limit is always finite, reflecting with the excitonic condensate regime in these systems. For the 3D semiconducting systems, the coherent DOS spectra survive also for the higher temperatures \cite{cite_20}. This situation is also typical for the double layer electronic structures at the half filling \cite{cite_21} when considering the electron-hole pair formation and condensation. In addition, it has been demonstrated that the excitonic insulator state and the excitonic condensation are two distinct phase transitions in the solid state \cite{cite_22, cite_23}, and the condensate states are due to the electronic phase stiffness \cite{cite_20,cite_24}, mechanism. Meanwhile, in the electronic bilayer systems, those mentioned phase transitions are indistinguishable as it was shown in Refs. \cite{cite_12, cite_21}. On the other hand, the inclusion of the dynamic screening and the full band structure favors the pairing and condensation \cite{cite_25, cite_26, cite_27}.  
 
In this paper, we use the bilayer Hubbard model to calculate the single-particle DOS functions in the BLG and we examine the excitonic effects in the DOS.
For the noninteracting layers, we show the principal modifications to the usual tight-binding single layer graphene's DOS behavior (with differences between the $A$ and $B$ DOS structures near the Dirac's neutrality point) and we estimate the excitonic blue-shift values in the normal DOS for the reasonable values of the interlayer hopping amplitude.

At finite values of the interlayer interaction parameter, the DOS shows always the remarkable four peaks structure and a very large interband hybridization gap opens for intermediate values of the interaction parameter. We have found the critical value of the interlayer interaction parameter at which the hybridization gap $\Delta_{\rm Hybr}$ opens in the system. We estimate the values of $\Delta_{\rm Hybr}$ for different strengths of the interlayer interaction parameter and for different values of the interlayer hopping amplitude $\gamma_1$. We show the modifications of the van Hove singularity (vHs.) peaks in the DOS when augmenting the interlayer interaction parameter. An interesting interlayer hopping mediated crossover, from the insulating pairing state to the excitonic condensate state follows from our considerations. Moreover, we show that at any reasonable value of the interlayer interaction parameter $W$, there exist a critical value of the interlayer hopping parameter $\gamma_1$ at which the hybridization gap vanishes and the BLG pass to the coherent excitonic condensate state. We calculate the hybridization gap in the system, and we show that it becomes larger when increasing the interaction parameter $W$. 

Furthermore, we estimate the excitonic shift energies, the intraband vHs. peaks separations and the values of the DOS at the Dirac's neutrality points for the zero temperature case. The full interaction bandwidth, considered here, mimics different limits of correlations in the BLG, which have been discussed only partially in the literature, where the discussions have been restricted only to the strong interlayer screening regime.

The paper is organized as follows: in the Section \ref{sec:Section_2}, we introduce the bilayer Hubbard model for our BLG system. In the Section \ref{sec:Section_3} we give the form of the fermionic action in the Feynman's path integral formalism. Next, in the Section \ref{sec:Section_4}, we discuss the single-particle DOS in the BLG, for different interlayer interaction regimes. In the Section \ref{sec:Section_5} we present the numerical results on the excitonic effects in the normal DOS and, in the Section \ref{sec:Section_6}, we give a short conclusion to our paper.

\section{\label{sec:Section_2} The interlayer Hubbard model}
%
The Bernal Stacked (BS) BLG system is composed of two coupled honeycomb layers with sublattice sites $A$, $B$ and $\tilde{A}$, $\tilde{B}$, in the bottom and top layers respectively, arranged in the $z$-direction in such a way that the atoms on the sites $\tilde{A}$ in the top layer lie just above the atoms on the sites $B$ in the bottom layer graphene, and each layer is composed of two interpenetrating triangular lattices.
We consider the electronic BLG structure with the equal chemical potentials and equal on-site quasienergies in each layer. When switching the local Coulomb potential between the layers, we keep the charge neutrality equilibrium across the BLG, by imposing the half-filling condition in each layer.
The interlayer Hubbard model with the intralayer $U$ and local interlayer $W$ Coulomb interaction terms is subjected by the following Hamiltonian 
\begin{eqnarray}
H&=&-\gamma_0\sum_{\left\langle {\bf{r}}{\bf{r}}'\right\rangle}\sum_{\sigma}\left(\bar{a}_{\sigma}({\bf{r}})b_{\sigma}({\bf{r}}')+h.c.\right)
\nonumber\\
&-&\gamma_0\sum_{\left\langle {\bf{r}}{\bf{r}}'\right\rangle}\sum_{\sigma}\left(\bar{\tilde{a}}_{\sigma}({\bf{r}})\tilde{b}_{\sigma}({\bf{r}}')+h.c.\right)
\nonumber\\
&-&\gamma_1\sum_{{\bf{r}}\sigma}\left(\bar{{b}}_{\sigma}({\bf{r}})\tilde{a}_{\sigma}({\bf{r}})+h.c.\right)-\sum_{{\bf{r}}\sigma}\sum_{\ell=1,2}\mu_{\ell}n_{\ell\sigma}({\bf{r}})
\nonumber\\
&+&U\sum_{{\bf{r}}}\sum_{\substack{\eta=ab \\ \tilde{a}\tilde{b}}}\left[\left(n_{\eta\uparrow}-1/2\right)\left(n_{\eta\downarrow}-1/2\right)-1/4\right]
\nonumber\\
&+&W\sum_{{\bf{r}}\sigma\sigma'}\left[\left(n_{1b\sigma}({\bf{r}})-1/2\right)\left(n_{2\tilde{a}\sigma'}({\bf{r}})-1/2\right)-1/4\right].
\nonumber\\
\label{Equation_1}
\end{eqnarray}
The first two terms describe the intralayer electron hopping with the hopping parameter ${\gamma_0}$. The summation $\left\langle {\bf{r}}{\bf{r}}' \right\rangle$, in these terms denotes the sum over the nearest neighbours lattice sites in the separated honeycomb layers in the BLG structure. We kept here the small letters $a,b$ and $\tilde{a}, \tilde{b}$ for the electron operators on the lattice sites $A,B$ and $\tilde{A},\tilde{B}$ respectively. The third term corresponds to the hopping between two different monolayers in the BLG and the parameter $\gamma_1$ is the interlayer hopping amplitude. We neglect the nonlocal interlayer hopping terms because their role is essentially unimportant in the considered problem about the excitonic effects. When working with the grand canonical ensemble, we have added also the chemical potential terms for each layer and for each sublattice. We suppose here that the chemical potentials $\mu_{\ell}$ corresponding to different layers $\ell=1,2$ and different sublattices $A,B$ and $\tilde{A},\tilde{B}$ are the same (this is, of course, true, if we consider purely electronic layers and we neglect the effect of the disorder, caused by the additionally charged impurities). It is important to mention here that the chemical potentials of electrons on the nonequivalent sublattice sites, in the same layer, get different shifts in different layers due to the stacking order of the BLG structure. Next, $n_{\ell\sigma}({\bf{r}})$ is the electron density operator for the fermions in the layer $\ell$ and with the spin $\sigma$. The four $U$-terms in the Hamiltonian in Eq.(\ref{Equation_1}) describe the on-site intralayer Coulomb interactions in the BLG. The summation index $\eta$ in the $U$-terms, in Eq.(\ref{Equation_1}), refers to different sublattice fermions in a given layer, i.e., $\eta=a, b$, for the bottom layer with $\ell=1$, and $\eta=\tilde{a}, \tilde{b}$, for the top layer with $\ell=2$.
We will study the excitonic effects in the BLG at the half-filling condition in each layer, i.e., $\left\langle n_{\ell} \right\rangle=1$, for $\ell=1,2$.
The last term in Eq.(\ref{Equation_1}), describes the local interlayer Coulomb repulsion, and the parameter $W$ is the corresponding interlayer Coulomb interaction.
We put $\gamma_{0}=1$, as the unit of energy, and we set $k_{B}=1$, $\hbar=1$ through the paper.
%
\section{\label{sec:Section_3} The fermionic action and Dirac representation}
%
Here, we write the partition function of our electronic BLG system. For simplifying the notations we will introduce the two component fermionic fields, corresponding to different type of fermions in the two sublattices of the graphene monolayers: $f_{c}=(\bar{c},c)$, where $c=a, b,$ for the layer with $\ell=1$ and $c=\tilde{a}, \tilde{b}$ for the layer with $\ell=2$. Next, the partition function will be written as 
\begin{eqnarray}
Z=\prod_{\sigma}\int\left[{\cal{D}}f_{a\sigma}{\cal{D}}f_{b\sigma}\right]\left[{\cal{D}}f_{\tilde{a}\sigma}{\cal{D}}f_{\tilde{b}\sigma}\right]e^{-{\cal{S}}\left[f_{a},f_{b},f_{\tilde{a}},f_{\tilde{b}}\right]}.
\label{Equation_2}
\end{eqnarray}
The fermionic action in the exponential, in Eq.(\ref{Equation_2}), is given by
\begin{eqnarray}
{\cal{S}}\left[f_{a},f_{b},f_{\tilde{a}},f_{\tilde{b}}\right]=\sum_{\substack{c=a,b \\ \tilde{a}, \tilde{b}}}{\cal{S}}_{\rm B}\left[f_c\right]+\int^{\beta}_{0}d\tau H\left(\tau\right).
\label{Equation_3}
\end{eqnarray}
We have introduced here the imaginary-time variables $\tau$ \cite{cite_28, cite_29}, at each lattice site ${\bf{r}}$ in both layers of the BLG. The variables $\tau$ vary in the interval $(0,\beta)$, where $\beta=1/T$ with $T$ being the temperature.
The first term in Eq.(\ref{Equation_3}) describes the fermionic Berry-terms, corresponding to all fermionic flavors in the BLG system. They are given as
\begin{eqnarray}
{\cal{S}}_{\rm B}\left[f_c\right]=\sum_{{\bf{r}}\sigma}\int^{\beta}_{0}d\tau \bar{c}_{\sigma}({\bf{r}}\tau)\frac{\partial}{\partial \tau}c_{\sigma}({\bf{r}}\tau).
\label{Equation_4}
\end{eqnarray}
Furthermore, in order to examine the excitonic effects in the BLG, we perform the real-space linearization of the four-fermionic terms, in Eq.(\ref{Equation_1}). We do not present here the details of such a procedure and we refer to our recent work, in Ref.\cite{cite_12}, where this procedure is presented in more details. 
For the next, we will consider the homogeneous BLG structure, when the pairing occurs between the particles with the same spin orientations, i.e., $\Delta_{\sigma\sigma'}=\Delta_{\sigma\sigma}\delta_{\sigma\sigma'}$ and we can assume that the pairing gap is real $\Delta_\sigma=\bar{\Delta}_{\sigma}\equiv\Delta$. Then the excitonic pairing gap parameter is given by 
\begin{eqnarray}
\Delta_{\sigma\sigma}=W\left\langle \bar{b}_{\sigma}({\bf{r}}\tau)\tilde{a}_{\sigma}({\bf{r}}\tau)\right\rangle.
\label{Equation_5}
\end{eqnarray}
Next, we pass to the Fourier space representation, given by the transformation $
c_{\sigma}({\bf{r}},\tau)=\frac{1}{\beta{N}}\sum_{{\bf{k}}\Omega_{n}}c_{\sigma{\bf{k}}}(\Omega_{n})e^{i\left({\bf{k}}{\bf{r}}-\Omega_{n}\tau\right)}$, where $N$ is the total number of sites on the $\eta$-type sublattice, in the layer $\ell$, and we write the partition function of the system in the form 

\begin{eqnarray} 
{\cal{S}}\left[\bar{\psi},\psi\right]=\frac{1}{\beta{{ N}}}\sum_{\substack{{\bf{k}}\Omega_n \\ \sigma}}\bar{\psi}_{\sigma{\bf{k}}}(\Omega_{n}){\hat{\cal{G}}}^{-1}_{\sigma{\bf{k}}}(\Omega_{n}){\psi}_{\sigma{\bf{k}}}(\Omega_{n}).
\label{Equation_6}
\end{eqnarray}
Here, $\Omega_{n}=\pi\left(2n+1\right)/\beta$ with $n=0,\pm1,\pm2,\dots$, are the fermionic Matsubara frequencies \cite{cite_29}. The four component Dirac spinors $\psi_{\sigma{\bf{k}}}(\Omega_n)$, in Eq.(\ref{Equation_6}), are introduced at each discrete state ${\bf{k}}$ in the reciprocal space and for each spin direction $\sigma=\uparrow, \downarrow$. Being the generalized Weyl spinors \cite{cite_30, cite_31}, they are given as

\begin{eqnarray} {\psi}_{\sigma{\bf{k}}}(\Omega_{n})=\left[a_{\sigma{\bf{k}}}(\Omega_{n}),b_{\sigma{\bf{k}}}(\Omega_{n}),\tilde{a}_{\sigma{\bf{k}}}(\Omega_{n}),\tilde{b}_{\sigma{\bf{k}}}(\Omega_{n})\right]^{T}.
\label{Equation_7}
\end{eqnarray}

The matrix ${\cal{G}}^{-1}_{\sigma{\bf{k}}}(\Omega_{n})$, in Eq.(\ref{Equation_6}), is the inverse Green's function matrix, of size $4\times4$. It is defined as
\begin{eqnarray}
\footnotesize
\arraycolsep=0pt
\medmuskip = 0mu
{\cal{G}}^{-1}_{\sigma{\bf{k}}}\left(\Omega_{n}\right)=\left(
\begin{array}{ccccrrrr}
E_{1}(\Omega_{n}) & -\tilde{\gamma}_{1{\bf{k}}} & 0 & 0\\
-\tilde{\gamma}^{\ast}_{1{\bf{k}}} &E_{2}(\Omega_{n})  & -\gamma_{1}-\bar{\Delta}_{\sigma} & 0 \\
0 & -\gamma_{1}-{\Delta}_{\sigma} & E_{2}(\Omega_{n}) & -\tilde{\gamma}_{2{\bf{k}}} \\
0 & 0 & -\tilde{\gamma}^{\ast}_{2{\bf{k}}} & E_{1}(\Omega_{n}) 
\end{array}
\right).
\label{Equation_8}
\end{eqnarray}
Indeed, the structure of the matrix does not changes when inverting the spin direction, i.e., ${G}^{-1}_{\downarrow{\bf{k}}}\left(\Omega_{n}\right)\equiv {G}^{-1}_{\uparrow{\bf{k}}}\left(\Omega_{n}\right)$.
The diagonal elements of the matrix, in Eq.(\ref{Equation_8}), are the quasienergies 
\begin{eqnarray}
E_{\ell}(\Omega_{n})=-i\Omega_{n}-\mu^{\rm eff}_{\ell},
\label{Equation_9}
\end{eqnarray}
where the effectve chemical potentials $\mu_{\ell}$ with $\ell=1,2$ are defined as $\mu^{\rm eff}_{1}=\mu+U/4$ and $\mu^{\rm eff}_{2}=\mu+U/4+W$. The parameters $\tilde{\gamma}_{\ell{\bf{k}}}$, in Eq.(\ref{Equation_8}), are the renormalized (nearest neighbors) intralayer hopping amplitudes: $\tilde{\gamma}_{\ell{\bf{k}}}=z\gamma_{\ell{\bf{k}}}\gamma_0$, where the ${\bf{k}}$-dependent parameters $\gamma_{\ell{\bf{k}}}$ are the energy dispersions in the BLG layers, Namely, we have 
\begin{eqnarray}
\gamma_{\ell{\bf{k}}}=\frac{1}{z}\sum_{{\vec{{\bf{\delta}}}}_{\ell}}e^{-i{{\bf{k}}{\vec{{\bf{\delta}}}}_{\ell}}},
\label{Equation_10}
\end{eqnarray}
The parameter $z$, is the number of the nearest neighbors lattice sites in the honeycomb lattice. The components of the nearest-neighbors vectors ${\vec{\bf{\delta}}}_{\ell}$, for the bottom layer 1, are given by ${\vec{\bf{\delta}}}^{(1)}_{1}=\left(d/2,d\sqrt{3}/2\right)$, ${\vec{\bf{\delta}}}^{(2)}_{1}=\left(d/2,-d\sqrt{3}/2\right)$ and ${\vec{\bf{\delta}}}^{(3)}_{1}=\left(-d,0\right)$. For the layer 2, we have obviously ${\vec{\bf{\delta}}}^{(1)}_{2}=\left(d,0\right)$, ${\vec{\bf{\delta}}}^{(2)}_{2}=\left(-d/2,-d\sqrt{3}/2\right)$, and ${\vec{\bf{\delta}}}^{(3)}_{2}=\left(-d/2,d\sqrt{3}/2\right)$. Then, for the function $\gamma_{1{\bf{k}}}$, we get
\begin{eqnarray}
\gamma_{1{\bf{k}}}=\frac{1}{3}\left(e^{-ik_{x}d}+2e^{i\frac{k_{x}d}{2}}\cos{\frac{\sqrt{3}}{2}k_{y}d}\right),
\label{Equation_11}
\end{eqnarray}
where $d$ is the carbon-carbon interatomic distance. It is not difficult to realize that $\gamma_{2{\bf{k}}}=\gamma^{\ast}_{1{\bf{k}}}\equiv\gamma^{\ast}_{{\bf{k}}}$, and we have $\tilde{\gamma}_{2{\bf{k}}}=\tilde{\gamma}^{\ast}_{1{\bf{k}}}\equiv\tilde{\gamma}^{\ast}_{{\bf{k}}}$,
where we have omitted the layer index $\ell$.   

The form of the Green's function matrix, given in Eq.(\ref{Equation_8}), has been used recently in the work of Ref.\cite{cite_12} in order to derive the self-consistent equations, which determine the excitonic gap parameter $\Delta$ and the effective bare chemical potential $\bar{\mu}$ in the interacting BLG system. Particularly, this last one plays an important role in the BLG theory and redefines the charge neutrality point (CNP) in the context of the exciton formation in the interacting BLG,\cite{cite_12}. Quite interesting experimental results on that subject are given recently in Refs.\cite{cite_32, cite_33, cite_34}.
%

\section{\label{sec:Section_4} The single-particle DOS}
%
\subsection{\label{sec:Section_4_1} The sublattice spectral functions}
\begin{figure}
	\begin{center}
		\includegraphics[width=240px,height=240px]{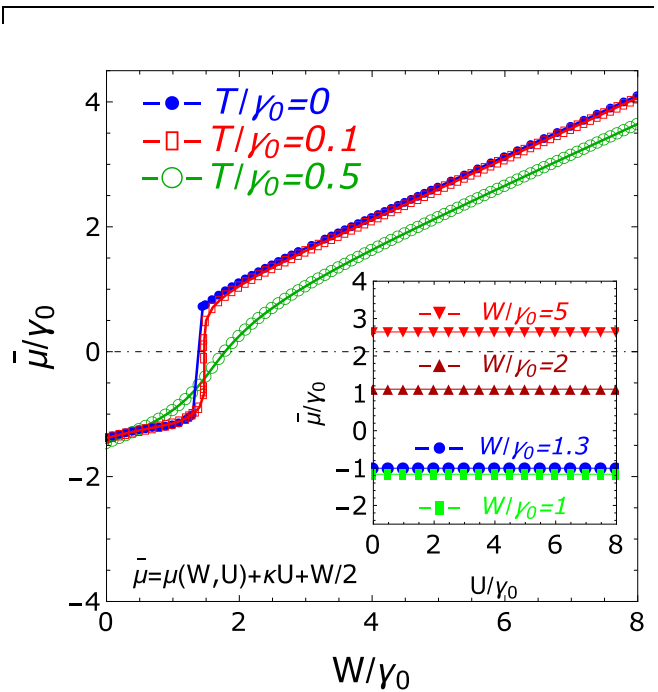}
		\caption{\label{fig:Fig_1}(Color online) The exact solution of the effective chemical potential, normalized to the intralayer hopping amplitude $\gamma_0$ as a function of the interlayer interaction parameter. The inset shows the dependence of $\bar{\eta}$ on the intralayer on-site interaction $U$, at $T=0$. The linear solution of the exact chemical potential $\mu$ is used in the calculations. The temperature is set at $T=0$. }
	\end{center}
\end{figure}

%
In this section, we calculate the single-particle normal DOS for the BLG system, given by the Hamiltonian, in Eq.(\ref{Equation_1}) and basing on the form of the fermionic action, given in Eq.(\ref{Equation_6}). The normal single-particle DOS is straightforwardly defined as 
\begin{eqnarray}
\rho_{c}(\omega)=\frac{1}{N}\sum_{{\bf{k}}}{\cal{S}}_{c}({\bf{k}},\omega),
\label{Equation_12}
\end{eqnarray}
where the index $c$ corresponds to the sublattice type in the BLG layers, i.e., $c=A,B,\tilde{A},\tilde{B}$, and the spectral function ${\cal{S}}_{c}({\bf{k}},\omega)$, in the right hand side in Eq.(\ref{Equation_12}), is defined with the help of the retarded real-time Green's function $G^{\rm R}_{c}({\bf{k}},\omega)$ \cite{cite_28, cite_29}
\begin{eqnarray}
{\cal{S}}_{c}({\bf{k}},\omega)=-\frac{1}{\pi}\Im{G^{\rm R}_{c}({\bf{k}},\omega)}.
\label{Equation_13}
\end{eqnarray} 
Here, we have supposed that the spin variable $\sigma$ is fixed in the direction spin-$\uparrow$, being completely unimportant for the considered here problem. In turn, the retarded Green function $G^{\rm R}_{c}({\bf{k}},\omega)$, could be obtained after the analytical continuation into the real frequency axis,
in the expression of the respective Matsubara Green's function
$G_{c}({\bf{k}},\Omega_{n})$, (see the similar procedures in Refs. \cite{cite_28, cite_29})
\begin{eqnarray}
G^{\rm R}_{c}({\bf{k}},\omega)=\left.G_{c}({\bf{k}},\Omega_{n})\right\vert_{i\Omega_{n}\rightarrow \omega+i0^{+}},
\label{Equation_14}
\end{eqnarray} 
The explicit calculation of the thermal normal Green's functions $G_{c}({\bf{k}},\Omega_{n})$ follows from the definition of the normal Matsubara Green's function \cite{cite_29}. As usually, in the real space, they are defined as the statistical average of the product of an annihilation $c$ and a creation type $\bar{c}$ operators, i.e., for our fermions, we have   
\begin{eqnarray}
G_{c}\left({\bf{r}}\tau,{\bf{r}}'\tau'\right)=-\left\langle c({\bf{r}}\tau)\bar{c}({\bf{r}}'\tau')\right\rangle.
\label{Equation_15}
\end{eqnarray}
After transforming into the Fourier space the fermionic operators, entering in Eq.(\ref{Equation_15}) the local expression of the Green's function (for the symmetry reasons of the action, in Eq.(\ref{Equation_6}), we consider the time- and space-local expression of the single particle Green's function), in Eq.(\ref{Equation_15}), is
\begin{eqnarray}
G_{c}\left({\bf{r}}\tau,{\bf{r}}\tau\right)=-\frac{1}{\beta{N}}\sum_{{\bf{k}}\Omega_{n}}G_{c}\left({\bf{k}},\Omega_{n}\right),
\label{Equation_16}
\end{eqnarray}
where the Fourier transforms $G_{c}\left({\bf{k}},\Omega_{n}\right)$ are given by
\begin{eqnarray}
G_{c}\left({\bf{k}},\Omega_{n}\right)=\frac{1}{\beta{N}}\left\langle c_{{\bf{k}}}(\Omega_{n})\bar{c}_{{\bf{k}}}(\Omega_{n})\right\rangle.
\label{Equation_17}
\end{eqnarray}
In order to calculate the statistical average on the right-hand side in Eq.(\ref{Equation_17}), we will perform the Hubbard-Stratanovich transformation in the expression of the partition function. In the Dirac's spinor notations, the partition function in Eq.(\ref{Equation_2}), in the Section \ref{sec:Section_3}, will be transformed as 
\begin{eqnarray}
Z=\int\left[{\cal{D}}\bar{\psi}{\cal{D}}\psi\right]e^{-\frac{1}{\beta{{ N}}}\sum_{\substack{{\bf{k}}\Omega_n \\ \sigma}}\bar{\psi}_{\sigma{\bf{k}}}(\Omega_{n}){\hat{\cal{G}}}^{-1}_{\sigma{\bf{k}}}(\Omega_{n}){\psi}_{\sigma{\bf{k}}}(\Omega_{n})}\times
\nonumber\\
\times 
e^{\frac{1}{\beta{{ N}}}\sum_{\substack{{\bf{k}}\Omega_n \\ \sigma}}\left[\frac{1}{2}\bar{J}_{{\bf{k}}\sigma}(\Omega_{n})\psi_{{\bf{k}}\sigma}(\Omega_{n})+\frac{1}{2}\bar{\psi}_{{\bf{k}}\sigma}(\Omega_{n}){J}_{{\bf{k}}\sigma}(\Omega_{n})\right]}\approx
\nonumber\\
\approx e^{\frac{1}{2}\sum_{\substack{{\bf{k}}\Omega_n \\ \sigma}}\bar{J}_{{\bf{k}}\sigma}(\Omega_{n}){\hat{\cal{G}}}_{\sigma{\bf{k}}}(\Omega_{n}){J}_{{\bf{k}}\sigma}(\Omega_{n})},
\label{Equation_18}
\end{eqnarray}
where the auxiliary source field vectors ${J}_{{\bf{k}}\sigma}(\Omega_{n})$ are the subjects of the Dirac's spinors defined similar to the $\psi$-fields, in the Section \ref{sec:Section_3}
\begin{eqnarray} J_{\sigma{\bf{k}}}(\Omega_{n})=\left[j_{a\sigma{\bf{k}}}(\Omega_{n}),j_{b\sigma{\bf{k}}}(\Omega_{n}),j_{\tilde{a}\sigma{\bf{k}}}(\Omega_{n}),j_{\tilde{b}\sigma{\bf{k}}}(\Omega_{n})\right]^{T}.
\label{Equation_19}
\end{eqnarray}
The matrix ${\hat{\cal{G}}}_{\sigma{\bf{k}}}(\Omega_{n})$ is defined as the inverse of the matrix given in Eq.(\ref{Equation_8}) and we have ${\hat{\cal{G}}}_{\sigma{\bf{k}}}(\Omega_{n})=\left[\frac{2}{\beta{N}}{\hat{\cal{G}}}^{-1}_{\sigma{\bf{k}}}(\Omega_{n})\right]^{-1}$.
Furthermore, the calculation of the thermal Matsubara Green's function, in Eq.(\ref{Equation_17}), is straightforward. For the sublattice-$A$, in the bottom layer of the BLG, we get
\begin{eqnarray}
\frac{\delta^{2}Z}{\delta{\bar{j}_{a\sigma{\bf{k}}}(\Omega_{n})}\delta{j_{a\sigma{\bf{k}}}(\Omega_{n})}}=-\frac{1}{4}\left\langle\bar{a}_{{\bf{k}}}(\Omega_{n})a_{{\bf{k}}}(\Omega_{n})\right\rangle=
\nonumber\\
=-\frac{\beta{N}}{4}{\cal{G}}^{11}_{{\bf{k}}}(\Omega_{n}).
\label{Equation_20}
\end{eqnarray}
Then it follows that
\begin{eqnarray}
{\cal{G}}^{11}_{{\bf{k}}}(\Omega_{n})=-\frac{1}{\beta{N}}\left\langle a_{{\bf{k}}}(\Omega_{n})\bar{a}_{{\bf{k}}}(\Omega_{n})\right\rangle.
\label{Equation_21}
\end{eqnarray}
After the inversion of the matrix given in Eq.(\ref{Equation_8}), the Green's function matrix component ${\cal{G}}^{11}_{{\bf{k}}}(\Omega_{n})$ takes the following form 
\begin{eqnarray}
{\cal{G}}^{11}_{{\bf{k}}}(\Omega_{n})=\sum^{4}_{i=1}\frac{\alpha_{i{\bf{k}}}}{i\Omega_{n}-\varepsilon_{i{\bf{k}}}},
\label{Equation_22}
\end{eqnarray}
where the dimensionless coefficients $\alpha^{(1)}_{i{\bf{k}}}$ are 
\begin{eqnarray}
\footnotesize
\arraycolsep=0pt
\medmuskip = 0mu
\alpha_{i{{\bf{k}}}}
=(-1)^{i+1}
\left\{
\begin{array}{cc}
\displaystyle  & \frac{{\cal{P}}^{(3)}(\varepsilon_{i{\bf{k}}})}{\left(\varepsilon_{1{\bf{k}}}-\varepsilon_{2{\bf{k}}}\right)}\prod^{}_{j=3,4}\frac{1}{\left(\varepsilon_{i{\bf{k}}}-\varepsilon_{j{\bf{k}}}\right)},  \ \ \  $if$ \ \ \ i=1,2,
\newline\\
\newline\\
\displaystyle  & \frac{{\cal{P}}^{(3)}(\varepsilon_{i{\bf{k}}})}{\left(\varepsilon_{3{\bf{k}}}-\varepsilon_{4{\bf{k}}}\right)}\prod^{}_{j=1,2}\frac{1}{\left(\varepsilon_{i{\bf{k}}}-\varepsilon_{j{\bf{k}}}\right)},  \ \ \  $if$ \ \ \ i=3,4,
\end{array}\right.
\nonumber\\
\label{Equation_23}
\end{eqnarray}
and ${\cal{P}}^{(3)}(\varepsilon_{i{\bf{k}}})$ is a polynomial of third order in $\varepsilon_{i{\bf{k}}}$. Namely, we have
\begin{eqnarray}	{\cal{P}}^{(3)}(\varepsilon_{i{\bf{k}}})=\varepsilon^{3}_{i{\bf{k}}}+\omega_{1{\bf{k}}}\varepsilon^{2}_{i{\bf{k}}}+\omega_{2{\bf{k}}}\varepsilon_{i{\bf{k}}}+\omega_{3\bf{k}}
\label{Equation_24}
\end{eqnarray}
with the coefficients $\omega_{i{\bf{k}}}$, $i=1,...3$, given as 
\begin{eqnarray}
&&\omega_{1{\bf{k}}}=-2\mu^{\rm eff}_{2}-\mu^{\rm eff}_{1},
\label{Equation_25} 
\newline\\
&&\omega_{2{\bf{k}}}=\mu^{\rm eff}_{2}\left(\mu^{\rm eff}_{2}+2\mu^{\rm eff}_{1}\right)-\left(\Delta+\gamma_1\right)^{2}-|\tilde{\gamma}_{{\bf{k}}}|^{2}, 
\label{Equation_26}
\end{eqnarray}
and
\begin{eqnarray}
&&\omega_{3{\bf{k}}}=-\mu^{\rm eff}_{1}\left(\mu^{\rm eff}_{2}\right)^{2}+\mu^{\rm eff}_{1}\left(\Delta+\gamma_1\right)^{2}+\mu^{\rm eff}_{2}|\tilde{\gamma}_{{\bf{k}}}|^{2}.
\label{Equation_27}
\end{eqnarray}
The quasiparticle energy parameters $\varepsilon_{i{\bf{k}}}$, in Eq.(\ref{Equation_23}) are defined as follows 
\begin{widetext}
\begin{eqnarray}
\varepsilon_{1,2{\bf{k}}}=\frac{1}{2}\left[\Delta+\gamma_{1}\pm\sqrt{\left(W-\Delta-\gamma_{1}\right)^{2}+4|\tilde{\gamma}_{{\bf{k}}}|^{2}}\right]-\bar{\mu},
\label{Equation_28}
\newline\\
\varepsilon_{3,4{\bf{k}}}=\frac{1}{2}\left[-\Delta-\gamma_{1}\pm\sqrt{\left(W+\Delta+\gamma_{1}\right)^{2}+4|\tilde{\gamma}_{{\bf{k}}}|^{2}}\right]-\bar{\mu},
\label{Equation_29}
\end{eqnarray}
\newline\\
\end{widetext}  
where we have introduced a new bare chemical potential 
\begin{eqnarray}
\bar{\mu}=\frac{\mu^{\rm eff}_{1}+\mu^{\rm eff}_{2}}{2}.
\label{Equation_30}
\end{eqnarray}
As we will show later on, the bare chemical potential has a fundamental impact on the DOS behavior in the BLG, and provide the excitonic shift on the frequency axis. 
The energy parameters in Eqs.(\ref{Equation_28}) and (\ref{Equation_29}) define the electronic band structure
in the BLG system with the excitonic pairing interaction (see also in Ref.(\cite{cite_12})). It is not diffucult to verify that for the noninteracting BLG, i.e., when $U=0$, $W=0$ and $\Delta=0$, the expressions in Eqs.(\ref{Equation_28}) and Eq.(\ref{Equation_29}) are reducing to the usual tight binding dispersion relations $\varepsilon_i=\pm\frac{\gamma_1}{2}\pm \sqrt{({k}\gamma_0)^{2}+(\gamma_1/2)^{2}}-\mu$, (with $k=|\gamma_{{\bf{k}}}|^{2}$), discussed in Ref.\cite{cite_35} in the context of the real-space Green's function study of the noninteractiong BLG. It is important to mention that the normal spectral functions in different layers of the BLG, coincide with each other when interchanging the sublattice notations in the monolayers. This follows from the symmetry of the action, given in Eq.(\ref{Equation_6}). Particularly, we obtain 
\begin{eqnarray}
{\cal{S}}_{\ell=2,c=\tilde{a}}({\bf{k}},\omega)&=&{\cal{S}}_{\ell=1,c=b}({\bf{k}},\omega),
\label{Equation_31}
\newline\\
{\cal{S}}_{\ell=2,c=\tilde{b}}({\bf{k}},\omega)&=&{\cal{S}}_{\ell=1, c=a}({\bf{k}},\omega).
\label{Equation_32}
\end{eqnarray}
In Fig.~\ref{fig:Fig_1}, we have presented the variation of the effective chemical potential $\bar{\mu}/\gamma_0$ as a function of the interlayer Coulomb interaction parameter $W/\gamma_0$. The temperature dependence has been also shown. In the inset, in Fig.~\ref{fig:Fig_1}, we have shown the dependence of $\bar{\mu}/\gamma_0$ on the local intralayer Coulomb interaction parameter $U$, given by 
\begin{eqnarray}
\bar{\mu}(W,U, \gamma_0, T)=\mu(W,U,\gamma_0, T)+\kappa{U}+\frac{W}{2},
\label{Equation_33}
\end{eqnarray}
where $\mu(W,U,\gamma_0, T)$ is the exact solution of the chemical potential in the BLG. Different fixed values of the interlayer interaction parameters $W$ are considered in the picture. The more detailed discussion on the role of the chemical potential is given in Ref.\cite{cite_12}.
We see that the behavior of the effective chemical potential as a function of $W/\gamma_0$ shows a finite, very large jump at $T/\gamma_0=0$ and the shifted CNP corresponds well with the recent experimental observations of the behavior of bilayer's average chemical potential, given in Ref.{\cite{cite_32}}, where a direct measurement of the chemical potential of BLG has been done, as a function of its carrier density $n$.
Here, it is important to mention that the excitonic effects in the BLG lead to the significant shift of the double CNP in the BLG (see in Refs.{\cite{cite_32, cite_33, cite_34}}). Contrary, the intensity of the function $\bar{\mu}/\gamma_0$ is much higher in our case, due to the single BLG considered here, while in the gated double BLG measurements, discussed in Ref.{\cite{cite_32}}, the interlayer interaction is much weaker as
compared with the double monolayer graphene (see about in Ref.{\cite{cite_10}}), and the reason
for this is the effect of the finite amount of carrier density $n_{T}$ induced in the top BLG Ref.{\cite{cite_32}}. 
%
%
\begin{figure}
	\begin{center}
		\includegraphics[width=240px,height=240px]{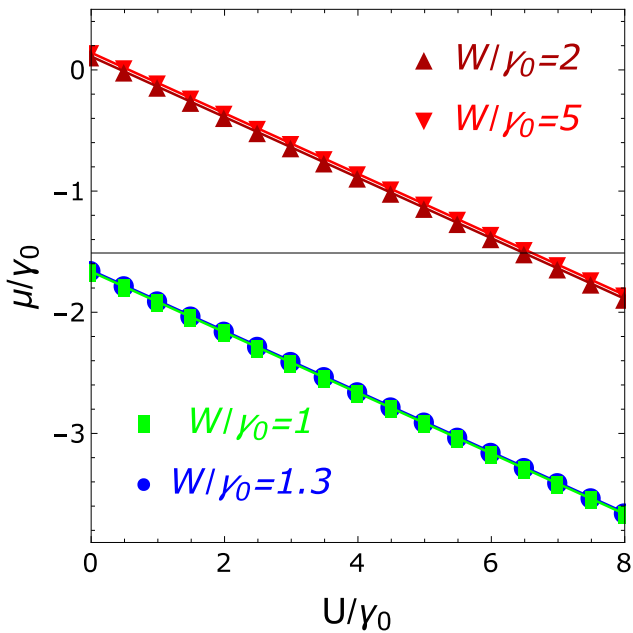}
		\caption{\label{fig:Fig_2}(Color online) The chemical potential solution of the BLG normalized to the intralayer hopping amplitude $\gamma_0$ as a function if the intralayer interaction parameter $U$. Different values of the interlayer interaction parameter $W$ are considered and the interlayer hopping amplitude is set at $\gamma_1=0.128\gamma_0$. The zero temperature limit is considered.}
	\end{center}
\end{figure} 
%
\begin{figure}
	\begin{center}
		\includegraphics[width=240px,height=310px]{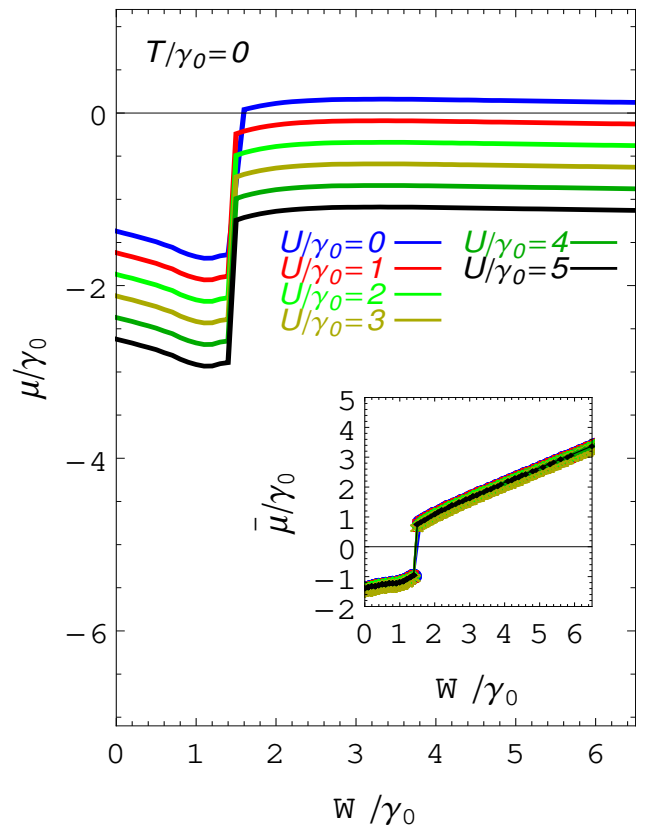}
		\caption{\label{fig:Fig_3}(Color online) The exact numerical solution for the chemical potential, normalized to the intralayer hopping amplitude $\gamma_0$, as a function of the interlayer Coulomb interaction parameter and for different values of the intralayer interaction parameter $U$. The inset: the effective chemical potential $\bar{\mu}$, calculated for the same values of the intralayer on-site interaction $U$ and given by the Eq.(\ref{Equation_33}).}
	\end{center}
\end{figure} 
%

In Fig.~\ref{fig:Fig_2}, the solution for the chemical potential is shown as a function of the intralayer Coulomb interaction parameter $U$ and for different values of the interlayer interaction parameter $W$. The linear slope of the chemical potential corresponds to the coefficient $\kappa=0.25$, given in Eq.(\ref{Equation_33}). In Fig.~\ref{fig:Fig_3}, the exact solution for $\mu$ is presented for different values of the intralayer interaction parameter $U$. The zero temperature case is considered in the picture and the interlayer hopping amplitude is fixed at the value $\gamma_1=0.128\gamma_0$.  
  
\subsection{\label{sec:Section_4_2} The sublattice density of states}
  
With the help of the analytical continuation, given in Eq.(\ref{Equation_14}) and by the use of the formula $1/(x+i\delta)={\cal{P}}(1/x)-i\pi\delta(x)$, we get for the single particle normal DOS for the sublattice $A$ 
\begin{eqnarray}
\rho_{A}(\omega)=\frac{1}{N}\sum_{{\bf{k}}}{\cal{S}}_{A}({\bf{k}},\omega)=
\nonumber\\
=\frac{1}{N}\sum^{4}_{i=1}\sum_{{\bf{k}}}\alpha_{i{\bf{k}}}\delta\left(\omega-\varepsilon_{i{\bf{k}}}\right).
\label{Equation_34}
\end{eqnarray}
Here, we can transform the summation over the wave vectors, into the integration over the continuous variables, by introducing the 2D density of states corresponding to the noninteracting graphene layers
\begin{eqnarray}
\rho_{\rm 2D}(x)=\frac{1}{N}\sum_{{\bf{k}}}\delta(x-\gamma_{{\bf{k}}}). 
\label{Equation_35}
\end{eqnarray}
The noninteracting DOS in the monolayer graphene beyond the Dirac's approximation \cite{cite_36, cite_37}, can be analytically expressed in terms of the elliptic integral of the first kind ${\textbf{K}}(x)$ \cite{cite_38}. Namely, we have
\begin{eqnarray}
\footnotesize
\arraycolsep=0pt
\medmuskip = 0mu\rho_{\rm 2D}(x)=\frac{2|x|}{\pi^{2}|\gamma_{0}|^{2}}
\left\{
\begin{array}{cc}
\displaystyle  & \frac{1}{\sqrt{\varphi\left(|{x}/{\gamma_0}|\right)}}{\textbf{K}}\left[\frac{4|x/\gamma_0|}{\varphi\left(|{x}/{\gamma_0}|\right)}\right],  \ \ \  0<|x|<\gamma_0,
\newline\\
\displaystyle  & \frac{1}{\sqrt{4|{{x}/{\gamma_0}}|}}{\textbf{K}}\left[\frac{\varphi\left(|x/\gamma_0|\right)}{4|{x}/{\gamma_0}|}\right],  \ \ \ \gamma_0<|x|<\infty,
\end{array}\right.
\nonumber\\
\label{Equation_36}
\end{eqnarray}
where, formally, we have enlarged the domain of variation of the argument, into $\pm\infty$. The function $\varphi(x)$, in Eq.(\ref{Equation_36}), is given by \cite{cite_36} 
\begin{eqnarray}
\varphi(x)=\left(1+x\right)^{2}-\frac{\left(x^{2}-1\right)^{2}}{4}.
\label{Equation_37}
\end{eqnarray}
After the definition in Eq.(\ref{Equation_37}), we get for the normal $A$ DOS function
\begin{eqnarray}
\rho_{A}(\omega)=\sum_{i,j=1,2}\frac{\rho_{\rm 2D}\left[\epsilon_i\left(\omega\right)\right]\alpha_{j}\left[\epsilon_i\left(\omega\right)\right]}{|\Lambda_{-}\left[\epsilon_1\left(\omega\right)\right]|}+
\nonumber\\
+\sum_{i,j=3,4}\frac{\rho_{\rm 2D}\left[\epsilon_i\left(\omega\right)\right]\alpha_{j}\left[\epsilon_i\left(\omega\right)\right]}{|\Lambda_{+}\left[\epsilon_3\left(\omega\right)\right]|},
\label{Equation_38}
\end{eqnarray}
where the frequency dependent dimensionless parameters $\epsilon_i(\omega)$ with $i=1,\dots, 4$, in Eq.(\ref{Equation_38}), are given by the following expressions
\begin{widetext}
\begin{eqnarray}
\footnotesize
\arraycolsep=0pt
\medmuskip = 0mu
&&\epsilon_i\left(\omega\right)=\frac{\left(-1\right)^{i+1}}{|z\gamma_0|}\left\{
\begin{array}{cc}
\displaystyle  & \sqrt{\left(\omega+\mu^{\rm eff}_{1}\right)\left(\omega+\mu^{\rm eff}_{2}-\gamma_{1}-\Delta\right)},  \ \ {\rm if}\ \   i=1,2,
\newline\\
\displaystyle  & {\sqrt{\left(\omega+\mu^{\rm eff}_{1}\right)\left(\omega+\mu^{\rm eff}_{2}+\gamma_{1}+\Delta\right)}},  \ \ {\rm if}\ \  i=3,4.
\end{array}\right.
\nonumber\\
\label{Equation_39}
\end{eqnarray}
\end{widetext}
Next, for the functions $\Lambda_{\mp}\left(x\right)$, in the denominators, in Eq.(\ref{Equation_38}), we have
\begin{eqnarray}
\Lambda_{\mp}\left(x\right)=\mp\frac{2|\gamma_0|^{2}x}{\sqrt{\left(W\mp\gamma_1\mp\Delta\right)^{2}+4|\gamma_0|^{2}x^{2}}}
\label{Equation_40}
\end{eqnarray}
and it is clear that $|\Lambda_{-}\left(\varepsilon_1\right)|=|\Lambda_{-}\left(\varepsilon_2\right)|$ and $|\Lambda_{+}\left(\varepsilon_3\right)|=|\Lambda_{+}\left(\varepsilon_4\right)|$. For the considered assumption of the half-filling in each layer, and as the theoretical and numerical calculations show, the normal single-particle DOS is not the same for different sublattices in the given layer with $l=1,2$ and near the shifted neutrality points. For the next, we will omit the layer indexes near the DOS functions notations (due to the relations in Eqs.(\ref{Equation_31}) and (\ref{Equation_32}), in the Section \ref{sec:Section_3}).
Similarly, for the sublattice $B$, we have the following expression for the $B$ DOS
\begin{eqnarray}
\rho_{B}(\omega)=\frac{1}{N}\sum_{{\bf{k}}}{\cal{S}}_{B}({\bf{k}},\omega)=
\nonumber\\
=\frac{1}{N}\sum^{4}_{i=1}\sum_{{\bf{k}}}\beta_{i{\bf{k}}}\delta\left(\omega-\varepsilon_{i{\bf{k}}}\right).
\label{Equation_41}
\end{eqnarray}
The coefficients $\beta_{i{{\bf{k}}}}$, in Eq.(\ref{Equation_41}) with $i=1,..4$ are given by the following relations
\begin{eqnarray}
\footnotesize
\arraycolsep=0pt
\medmuskip = 0mu
\beta_{i{{\bf{k}}}}=(-1)^{i+1}
\left\{
\begin{array}{cc}
\displaystyle  & \frac{{\cal{P}'}^{(3)}(\varepsilon_{i{\bf{k}}})}{\left(\varepsilon_{1{\bf{k}}}-\varepsilon_{2{\bf{k}}}\right)}\prod^{}_{j=3,4}\frac{1}{\left(\varepsilon_{i{\bf{k}}}-\varepsilon_{j{\bf{k}}}\right)},  \ \ \  $if$ \ \ \ i=1,2,
\newline\\
\newline\\
\displaystyle  & \frac{{\cal{P}'}^{(3)}(\varepsilon_{i{\bf{k}}})}{\left(\varepsilon_{3{\bf{k}}}-\varepsilon_{4{\bf{k}}}\right)}\prod^{}_{j=1,2}\frac{1}{\left(\varepsilon_{i{\bf{k}}}-\varepsilon_{j{\bf{k}}}\right)},  \ \ \  $if$ \ \ \ i=3,4,
\end{array}\right.
\nonumber\\
\label{Equation_42}
\end{eqnarray}
where ${\cal{P}'}^{(3)}(\varepsilon_{i{\bf{k}}})$ in Eq.(\ref{Equation_42}) is again a polynomial of third order in $\varepsilon_{i{\bf{k}}}$, namely

\begin{eqnarray}	{\cal{P}'}^{(3)}(\varepsilon_{i{\bf{k}}})=\varepsilon^{3}_{i{\bf{k}}}+\omega'_{1{\bf{k}}}\varepsilon^{2}_{i{\bf{k}}}+\omega'_{2{\bf{k}}}\varepsilon_{i{\bf{k}}}+\omega'_{3\bf{k}}
\label{Equation_43}
\end{eqnarray}
with the coefficients $\omega'_{i{\bf{k}}}$, $i=1,...3$, given as 
\begin{eqnarray}
&&\omega'_{1{\bf{k}}}=-2\mu^{\rm eff}_{1}-\mu^{\rm eff}_{2},
\label{Equation_44} 
\newline\\
&&\omega'_{2{\bf{k}}}=\mu^{\rm eff}_{1}\left(\mu^{\rm eff}_{1}+2\mu^{\rm eff}_{2}\right)-|\tilde{\gamma}_{{\bf{k}}}|^{2},
\label{Equation_45}
\end{eqnarray}
and
\begin{eqnarray}
&&\omega'_{3{\bf{k}}}=-\mu^{\rm eff}_{2}\left(\mu^{\rm eff}_{1}\right)^{2}+\mu^{\rm eff}_{1}|\tilde{\gamma}_{{\bf{k}}}|^{2}.
\label{Equation_46}
\end{eqnarray}
We see that the coefficients $\omega'_{1{\bf{k}}}$, $\omega'_{2{\bf{k}}}$ and $\omega'_{3{\bf{k}}}$ could be obtained from the coefficients $\omega_{1{\bf{k}}}$, $\omega_{2{\bf{k}}}$ and $\omega_{3{\bf{k}}}$, just by replacing the effective chemical potentials $\mu^{(1)}_{\rm eff}\rightleftharpoons \mu^{(2)}_{\rm eff}$ and by setting simultaneously $\Delta=0$.
Finally, for the single-particle $B$ DOS we get
\begin{eqnarray}
\rho_{B}(\omega)=\sum_{i,j=1,2}\frac{\rho_{\rm 2D}\left[\epsilon_i\left(\omega\right)\right]\beta_{j}\left[\epsilon_i\left(\omega\right)\right]}{|\Lambda_{-}\left[\epsilon_1\left(\omega\right)\right]|}+
\nonumber\\
+\sum_{i,j=3,4}\frac{\rho_{\rm 2D}\left[\epsilon_i\left(\omega\right)\right]\beta_{j}\left[\epsilon_i\left(\omega\right)\right]}{|\Lambda_{+}\left[\epsilon_3\left(\omega\right)\right]|}.
\label{Equation_47}
\end{eqnarray} 
We will examine numerically calculated DOS functions in the next Section of the present paper. 
%
\section{\label{sec:Section_5} Results and Discussion}
\begin{figure}
\begin{center}
\includegraphics[width=250px,height=470px]{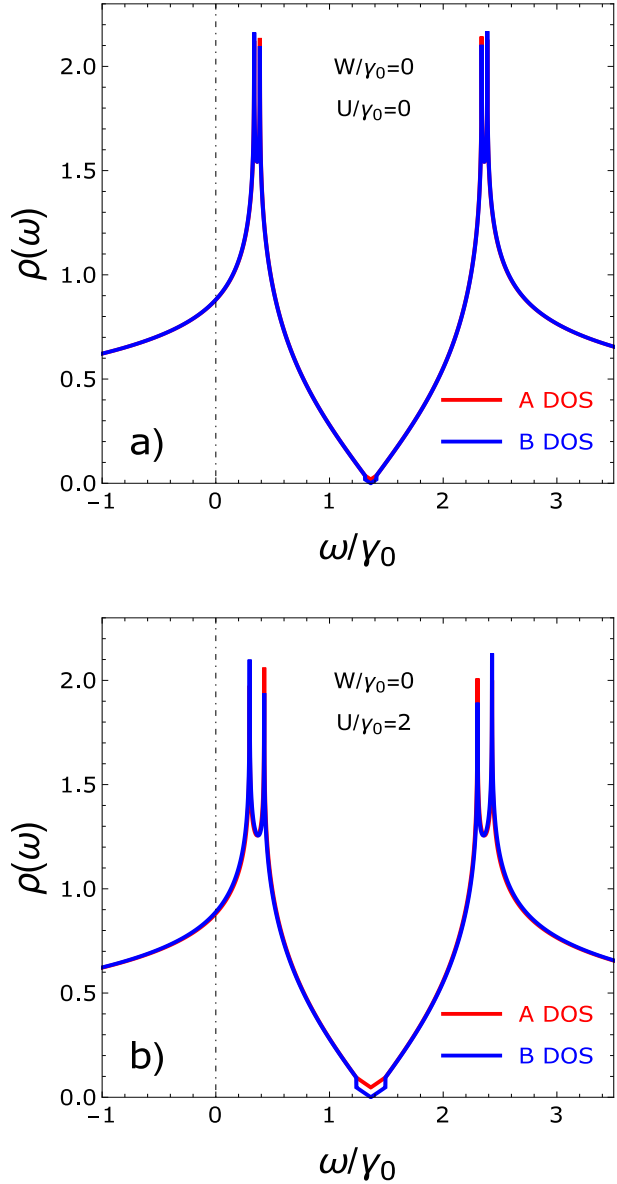}
\caption{\label{fig:Fig_4}(Color online) The $A$ and $B$ sublattice DOS functions at the zero interlayer Coulomb interaction. In the upper panel (a) the DOS functions are shown for the zero intralayer interaction $U$. In the lower panel (b) the same functions are shown for a finite value of the interaction parameter $U=2\gamma_0$.}
\end{center}
\end{figure}

\begin{figure}
	\begin{center}
		\includegraphics[width=210px,height=500px]{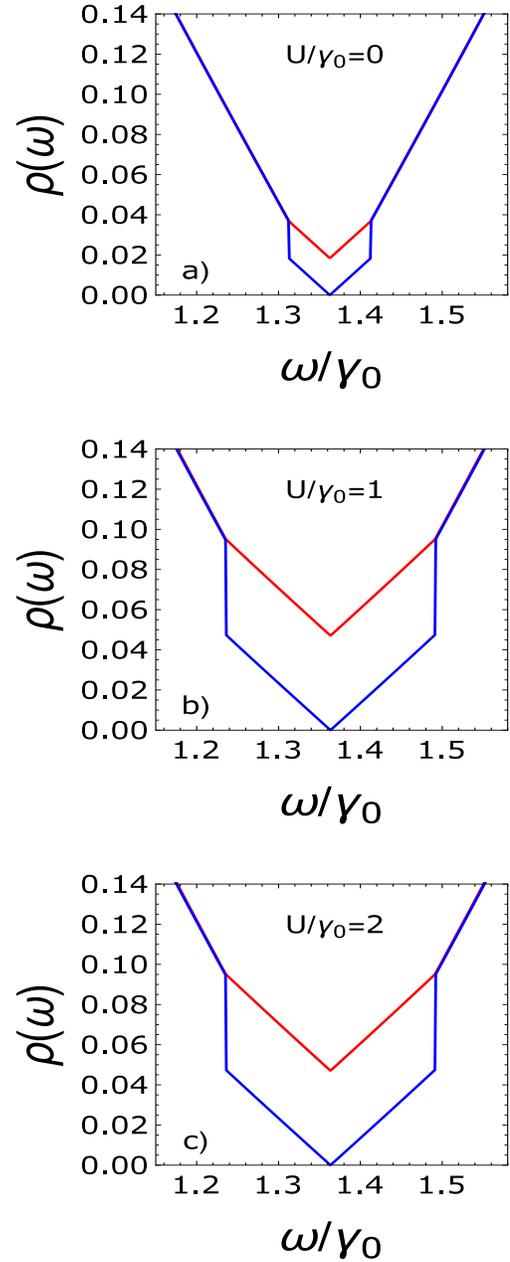}
		\caption{\label{fig:Fig_5}(Color online) The $A$ and $B$ sublattice DOS functions evolutions as a function of the interaction parameter $U$ (from (a) to (c), in the picture). The zero interlayer coupling case is considered and the interlayer hopping amplitude is fixed at $\gamma_1=0.128\gamma_0$. The zero temperature case is considered.}
	\end{center}
\end{figure}

%
In the panels a and b, in Fig.~\ref{fig:Fig_4}, we have presented the plots of the $A$ and $B$ DOS functions given in Eqs.(\ref{Equation_38}) and (\ref{Equation_47}). The zero interlayer interaction limit is considered $W=0$. In the panel a, in Fig.~\ref{fig:Fig_4}, the plots of the $A$ and $B$ DOS functions are presented for the case of the zero intralayer Coulomb interaction $U$. It is clear in Fig.~\ref{fig:Fig_4} that each band of the BLG, given in Eqs.(\ref{Equation_28}) and Eq.(\ref{Equation_29}), contributes with one vHs. peak inherited from the single-layer graphene spectrum. We see that the DOS functions, given in Eqs.(\ref{Equation_38}) and (\ref{Equation_47}) reproduce correctly the tight binding graphene DOS behavior (see in Ref.{\cite{cite_36}}) with the difference that the neutrality Dirac's point is shifted toward the higher frequencies, $\omega_0=1.363\gamma_0$. For the realistic $\gamma_0=3 eV$ (see in Ref.{\cite{cite_39}}), we get for the shifted frequency $\omega_0=4.089$ eV, which is sufficiently large as compared with the tight binding graphene's value. In the panel b, in Fig.~\ref{fig:Fig_4}, the same DOS functions are plotted for the case of the finite intralayer interaction parameter $U=2\gamma_0$. As we see, in this case, the separation between the vHs. peaks in the DOS is larger, and the value of the DOS at the neutrality point $\omega_0$, for the sublattice $A$, is higher than in the previous case, given in the panel a. The position of the neutrality point is unchanged, and we see also that the region, where the $B$ DOS is drastically decreasing, is much larger in the case of the nonzero intralayer interaction parameter $U\neq0$. The difference between the $A$ and $B$ DOS structures is clear in the panel b in Fig.~\ref{fig:Fig_4}. In Fig.~\ref{fig:Fig_5}, we have presented the DOS functions for different values of the intralayer Coulomb interaction parameter: $U=0$, $U=1\gamma_0$ and $U=2\gamma_0$. In all that cases the DOS functions, corresponding to different sublattices, are different near the neutrality point. It is important to mention that the DOS functions remain unchanged when $U\neq 0$, but they are different from the zero interaction case $U=0$. The DOS behaviors presented in Figs.~\ref{fig:Fig_4} and ~\ref{fig:Fig_5} are very similar to the DOS structures, discussed in Ref.\cite{cite_35}, apart the shifted neutrality point. The shift effect of the neutrality point in the DOS structures is due to the strong excitonic effects in the BLG.  

\begin{widetext}
	
	\begin{figure}
		\begin{center}
			\includegraphics[width=500px,height=170px]{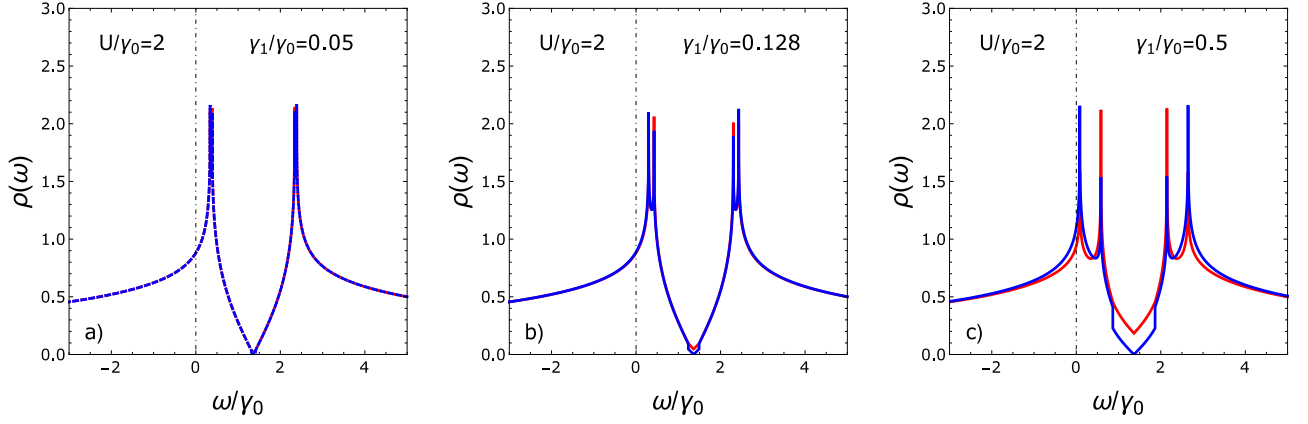}
			\caption{\label{fig:Fig_6}(Color online) The evolution of the single-particle DOS functions for different values (from (a) to (c)) of the interlayer hopping amplitude $\gamma_1$. The zero interlayer interaction and the zero temperature cases are considered. The formation of the coherent excitonic condensate states is shown in the panel (c). }
		\end{center}
	\end{figure} 
	
\end{widetext}

\begin{figure}
	\begin{center}
		\includegraphics[width=230px,height=220px]{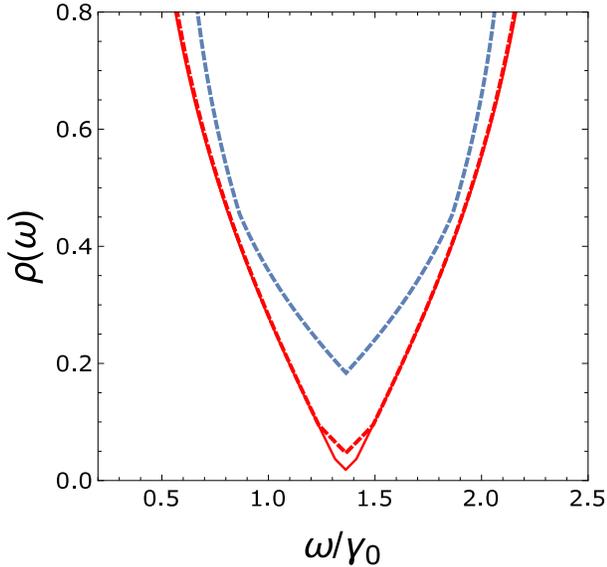}
		\caption{\label{fig:Fig_7}(Color online) The $A$ sublattice DOS function evolution near the Dirac's neutrality point $\omega_0$ for different values of the interlayer hopping amplitude $\gamma_1$. The zero interlayer interaction is considered. The values $\gamma_1=0.05\gamma_0$ (red solid line), $\gamma_1=0.128\gamma_0$ (red dotted line) and $\gamma_1=0.5\gamma_0$ (cerulean dotted line) are considered in the picture.}
	\end{center}
\end{figure}

In Fig.~\ref{fig:Fig_6}, we have shown the the $A$ and $B$ DOS evolutions for different values of the interlayer hopping amplitude $\gamma_1$. The intralayer Coulomb interaction parameter is fixed at the value $U=2\gamma_0$, and the zero interlayer Coulomb interaction case is considered in the picture. We observe in Fig.~\ref{fig:Fig_6} (see in the panels b and c) that the increase of the interlayer hopping amplitude leads to a very large number of $A$ DOS at the neutrality point. The vHs. peaks separations also become very large when increasing the parameter $\gamma_1$. In Fig.~\ref{fig:Fig_7}, we have presented the evolution of the $A$ DOS near the neutrality point for the same values of the interlayer hopping amplitude $\gamma_1$. The $A$ DOS behavior near the point $\omega_0$ shows that the interlayer hopping amplitude could lead to the existence of the interlayer excitonic condensate states even at the zero value of the interlayer Coulomb interaction parameter. The very large $A$ DOS value at the Dirac's point (see the panel c, in Fig.~\ref{fig:Fig_6}, and also in Fig.~\ref{fig:Fig_7}) is caused by the shift of higher situated energy states in the $A$ DOS structure toward the neutrality point $\omega_0$, and mediated by the formation of coherent condensates states. This scenario of the excitonic condensation at the zero interlayer coupling is converging well with the general discussion about coherent excitonic density of states in the semiconducting systems (see in Ref.\cite{cite_20}), where it has been shown that a large amount of states in the DOS (without the hybridization gap) is the sign of the coherent excitonic condensates in these systems.      
%
\subsection{\label{sec:Section_5_1} The hybridization gap in the BLG DOS}
\begin{figure}
	\begin{center}
		\includegraphics[width=240px,height=430px]{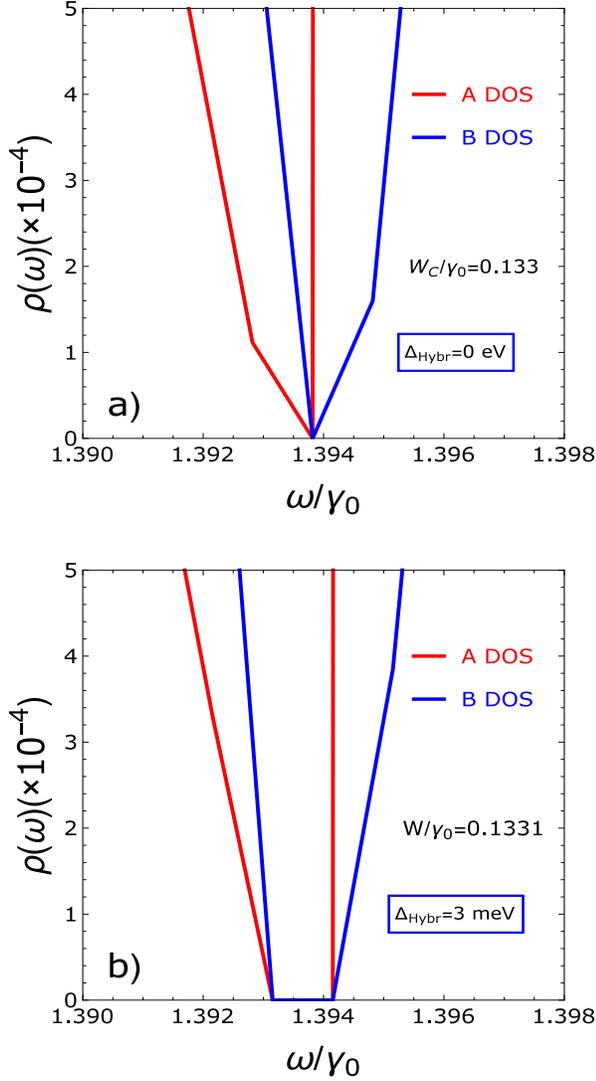}
		\caption{\label{fig:Fig_8}(Color online) Panel a: the DOS functions at the critical value of the interlayer interaction parameter $W_c=0.133\gamma_0$. Panel b: the formation of the symmetric hybridization gap in the DOS spectrum above the critical value $W_c$: $W=1.00075 W_c$. The interlayer hopping amplitude is fixed at $\gamma_1=0.128\gamma_0$ for both panels.}
	\end{center}
\end{figure}

\begin{figure}
	\begin{center}
		\includegraphics[width=240px,height=240px]{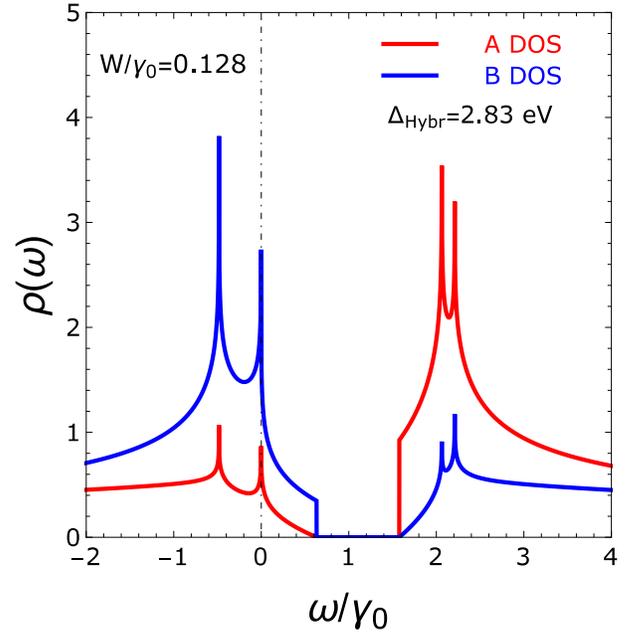}
		\caption{\label{fig:Fig_9}(Color online) The hybridization gap formation at $W=1.26\gamma_0=3.78$ eV (corresponding to the maximum value of the excitonic pairing gap parameter $\Delta=0.1867\gamma_0=0.56$ eV, discussed in Ref.{\cite{cite_12})} and for the interlayer hopping amplitude fixed at $\gamma_1=0.128\gamma_0=0.384$ eV. The zero temperature case is considered in the picture.}
	\end{center}
\end{figure}

%
Here, we will examine the formation of the hybridization gap in the BLG system caused by the interlayer Coulomb interaction parameter $W$. For the convenience, we will fix the value of the interlayer hopping amplitude at the value $\gamma_1=0.128\gamma_0$ and the intralayer interaction parameter at $U=2\gamma_0$. In  Fig.~\ref{fig:Fig_8}, we have presented the enlarged pictures of the $A$ and $B$ sublattice DOS functions near the neutrality point $\omega_0$. In Fig.~\ref{fig:Fig_8}, we consider two, very close, values of the interlayer interaction parameter $W$ and we show how the hybridization gap appear above the critical value $W=0.133\gamma_0$ (for a given value of $\gamma_1$). We see in the upper panel a, in Fig.~\ref{fig:Fig_8}, that the behaviors of different sublattice DOS functions are drastically not the same near the neutrality Dirac's point $\omega_0$ in the DOS. As the numerical calculations show, there is a critical value of the interlayer interaction parameter, above which the hybridization gap starts to open in the BLG (see in the lower panel b in Fig.~\ref{fig:Fig_8}) and the bilayer system passes to the regime of the excitonic pairing in the insulating state. A very small, but finite hybridization gap appears at $W=0.1331\gamma_0$, which is slightly higher than the critical value $W=0.133\gamma_0$, considered in the upper panel a, in Fig.~\ref{fig:Fig_8}. In Fig.~\ref{fig:Fig_9}, we have shown the $A$ and $B$ sublattice DOS functions for the large value of the interlayer Coulomb interaction parameter, corresponding to the maximum value of the excitonic pairing gap parameter (see in Ref.\cite{cite_12}). It is clear from the structure of the DOS functions, presented in Fig.~\ref{fig:Fig_9}, that the insulating state of the BLG is symmetric with respect to the hybridization gap formation, i.e., the $A$ and $B$ DOS functions goes to zero at the same values on frequency axes on the both sides of the hybridization gap $\Delta_{\rm Hybr}$.      
It is remarkable to note also that unlike the half-filling considered here, the vHs. in the DOS structures, corresponding to different particle channels in the band structure, are not symmetric with respect to the Dirac's point and the hybridization gap. The inter-peak separations in the particle or hole channel in the DOS become strongly asymmetric for the finite values of the interlayer interaction parameter $W$.
The observed "blue" shift effect of the neutrality point and the strong asymmetries in the DOS structures are due to the strong excitonic effects in the BLG. At any finite value of the interlayer interaction parameter, the BLG system is in the excitonic pairing state. On the other hand, the excitonic condensation is impossible for the large values of $W$ (even at large values of the parameter $\gamma_1$), because of the strongly hybridized states in the particle and hole channels and the very large hybridization gap.

%
%

\begin{figure}
	\begin{center}
		\includegraphics[width=240px,height=230px]{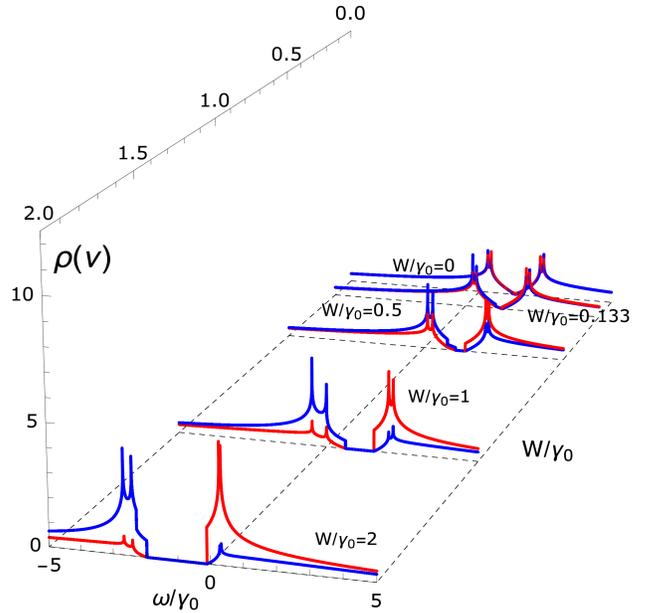}
		\caption{\label{fig:Fig_10}(Color online) The evolution of the $A$ and $B$ sublattice DOS functions for different values of the interlayer interaction parameter $W$ (see the values $W=0$, $W=0.133\gamma_0$, $W=0.5\gamma_0$, $W=1\gamma_0$ and $W=2\gamma_0$ in the picture). The interlayer hopping amplitude is set at $\gamma_1=0.128\gamma_0$, and the zero temperature limit is considered. }
	\end{center}
\end{figure}

In Fig.~\ref{fig:Fig_10}, we have shown the $W$ dependence of the $A$ and $B$ sublattice DOS functions for the interlayer hopping amplitude $\gamma_1=0.128\gamma_0$ and for $U=2\gamma_0$. Five values of the interaction parameter are indicated in Fig.~\ref{fig:Fig_10}. Just for the interest, we have shown also the DOS functions at the critical value of the interaction parameter $W_c=0.133\gamma_0=0.399$ eV, above which the hybridization gap opens in the BLG. The principal observation in Fig.~\ref{fig:Fig_10} is that the hybridization gap is increasing when increasing the interlayer interaction parameter. This fact is in good agreement with the principal results of the excitonic pairing scenario discussed in Ref.\cite{cite_12}, where it has been shown that the interlayer coupling interaction favors the excitonic pairing state in the BLG. 
%
%
%

\begin{figure}
	\begin{center}
		\includegraphics[width=240px,height=230px]{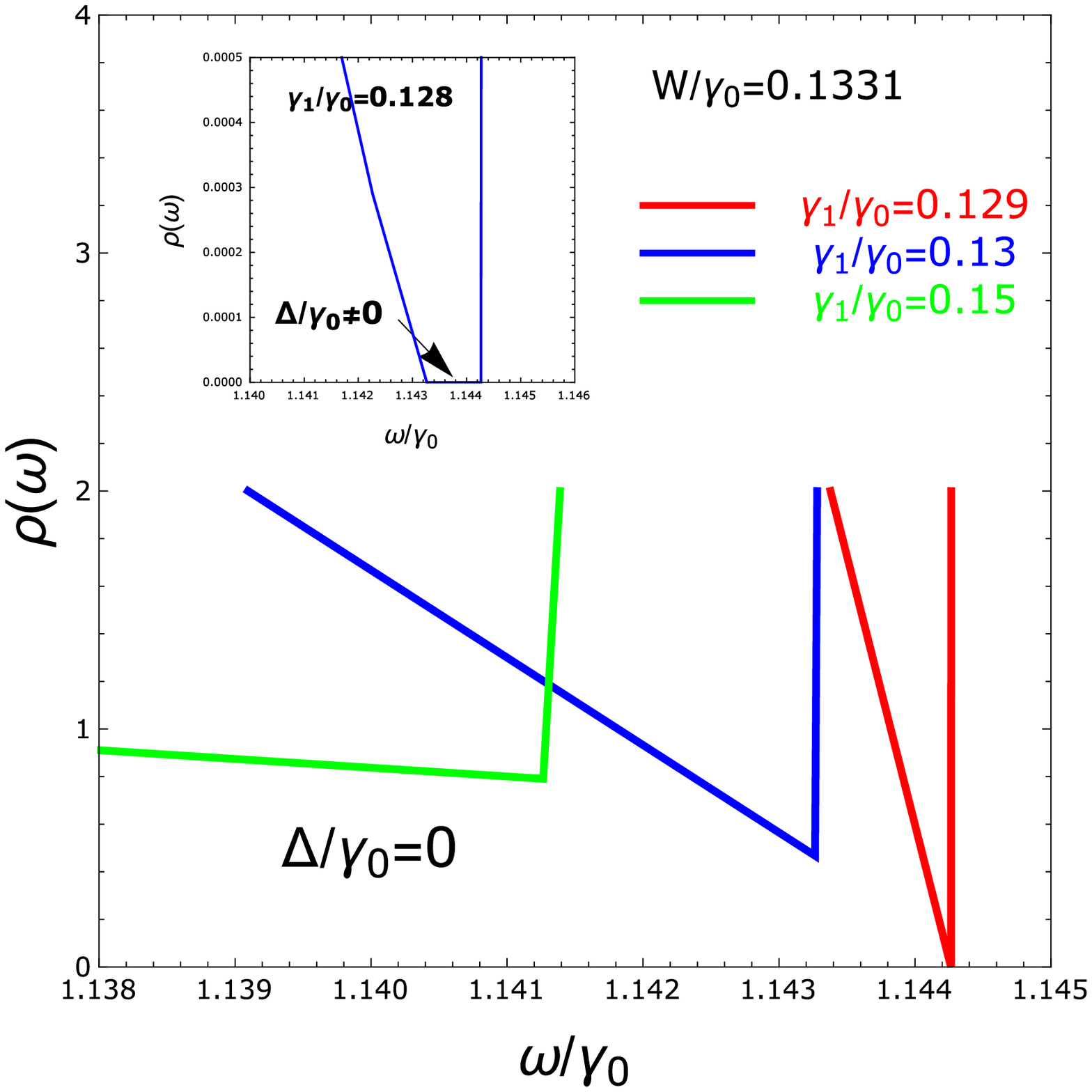}
		\caption{\label{fig:Fig_11}(Color online) The opening of the hybridization gap at $T=0$. The interlayer interaction parameter is fixed at $W=0.1331\gamma_0$. Different values of the interlayer hopping amplitude are considered.}
	\end{center}
\end{figure}

%
On the other hand, in Fig.~\ref{fig:Fig_11}, we have shown how the insulating gap, in the $A$ DOS spectrum, is closing when augmenting the interlayer hopping amplitude (the relatively small interlayer interaction parameter is considered in Fig.~\ref{fig:Fig_11}: $W=0.1331\gamma_0$). In the inset, in Fig.~\ref{fig:Fig_11}, we have shown the $A$ DOS spectrum with the hybridization gap of order $\Delta_{\rm Hybr}=0.00299\gamma_0=8.97$ meV. The interlayer Coulomb interaction parameter is of order $W=0.1331\gamma_0=399$ meV and the interlayer hopping amplitude is $\gamma_1=0.128\gamma_0=384$ meV. Then, in the picture in Fig.~\ref{fig:Fig_11}, we show the $A$ DOS near the Dirac's point and for three different values of the interlayer hopping amplitude $\gamma_1=0.129\gamma_0$,  $\gamma_1=0.13\gamma_0$ and  $\gamma_1=0.15\gamma_0$ (from right to the left). At the very large value of the parameter $\gamma_1$ there is a large number of $A$ DOS at the neutrality point (see the green line in Fig.~\ref{fig:Fig_11}), which could correspond to the formation of the interlayer excitonic condensate states even at the non-zero value of the interlayer interaction parameter $W$. Thus at the large values of the interlayer hopping amplitude, the system BLG is passing from the insulating hybridized state into the possible excitonic condensate state. This improvement analog to this, about the excitonic condensate state in the BLG and mediated by the parameter $\gamma_1$ is also discussed in Ref.{\cite{cite_12}}, where it has been shown how the excitonic condensation state is improved for the large interlayer hoppings. 
%
%
%
Let's mention also that the value of the excitonic shift-frequency $\omega_{0}=4.089$ eV is very close to the absolute numerical value of the effective bare chemical potential solution in the BLG: $|\bar{\mu}|=1.37\gamma_0=4.11$ eV, and which has been calculated in Ref.\cite{cite_12}. The important role of the non-zero chemical potential solution on the DOS behavior is discussed also in Ref.\cite{cite_40}, concerning the single layer graphene, where a Drude peak arises in the longitudinal conductivity spectrum, and the DOS becomes finite at the Fermi level. We observe also in Fig.~\ref{fig:Fig_11}, that the Dirac's neutrality point $\omega_0$ is shifting toward the lower frequency region (see the evolution from red to green lines in the picture). This \textbf{red-shift } effect and the excitonic shift observed in the previous pictures, presented here, are much more significant than the shift effects discussed in Refs.\cite{cite_36, cite_40}, which are due to the inclusion of the next nearest neighbors intra- and interlayer hoppings in the monolayer graphene and BLG, and also differ from the results on the single impurity problem, discussed in Ref.\cite{cite_41}. The increase of the interlayer interaction parameter above its critical value $W_c$, with the appropriate highest value of the interlayer hopping, (see the green line, in Fig.~\ref{fig:Fig_11}) leads to the right shift of the Dirac's frequency $\omega_0$ in the $A$ DOS with 
$\omega_0=1.14126\gamma_0=3.423$ eV for $W=0.1331\gamma_0=0.3993$ eV and $\gamma_1=0.15\gamma_0=0.45$ eV, in comparison with $\omega_0=1.393\gamma_0=4.179$ eV and $\gamma_1=0.128\gamma_0=0.384$ eV, corresponding to the critical value $W_c=0.133\gamma_0=0.399$ eV.     
%
%
%
It is remarkable to note also that the large values of A-DOS at the $\omega_0$ shift-points could correspond in this case also to the finite excitonic quasiparticle lifetimes at the given condensate state of the BLG, apart from its artifact-significance as the sublattice DOS. 
This effect of the dependence of the $A$ DOS on the interlayer hopping amplitude and the possible formation of the interlayer condensate states will undoubtedly have its impact on the excitonic absorption spectrum in the BLG both at zero interlayer coupling (zero applied bias) and nonzero coupling cases and should be verified ulteriorly (maybe the citations are needed). Particularly, we expect that in a large domain of the incident photon's energies, the BLG absorption spectrum, at the zero applied voltage, will show a sufficiently large absorption peak region in the case of the large interlayer hopping amplitude $\gamma_1$. In contrast, for the finite interlayer Coulomb interaction, the $A$ DOS has finite values at the neutrality points in the case of the relatively small Coulomb interactions $W$ and relatively high values of the interlayer hopping $\gamma_1$. This blue-shift effect, caused by the interlayer hopping amplitude, means that the excitonic condensate states survive for the higher values of the incident photon's energies, thus improving the excitonic insulator state at the large values interlayer hopping.
%
\section{\label{sec:Section_6} Conclusion}
%

We have considered the density of states in the BLG system, by considering the bilayer Hubbard model at the half-filling condition in each layer, and by assuming the statistical equilibrium states for each value of the interlayer interaction parameter. The theoretical method considered here permits to obtain the important results for the effective chemical potential in the BLG, which shows the extraordinary close results with the recent experimental measurements of the chemical potential in the gated BLG and double BLG heterostructures. For the first time in the literature, we show theoretically, how the charge neutrality point is changing its position when considering the excitonic effects in the BLG system. 

We have calculated the $A$ and $B$ sublattice DOS functions in the BLG for different interlayer interaction regimes and for different values of the interlayer hopping parameter. At the zero interlayer interaction case, we have obtained the results very similar to the usual tight-binding DOS in the BLG, and a very large "blue"-shift of the Dirac's neutrality point mediated by the strong excitonic effects in the BLG. At the zero interlayer coupling limit, we have shown the main modifications to the usual tight-binding DOS. We have shown that the excitonic condensation mechanism in the charge equilibrated BLG is possible even in the case of the noninteracting layers of the BLG. The principal tunable parameter, in this case, is the interlayer hopping amplitude $\gamma_1$, the large values of which improves the excitonic condensate state. In addition, at any finite and realistic value of the interlayer interaction parameter, it is possible to find the critical value of the interlayer hopping amplitude that renders the BLG into the excitonic condensation state just by suppressing the hybridization gap, present in the system. For example, at the value, $W=0.1331\gamma_0$ and at $\gamma_1=0.128\gamma_0$ a very small but finite hybridization gap is present in the BLG. When slightly augmenting the parameter $\gamma_1$ to $\gamma_1=0.129\gamma_0$, the hybridization gap is suppressed and the system starts to pass into the excitonic condensate regime. The insulating excitonic state is also suppressed in this case. The principal consequence from this consideration is the following statement: \textbf{Statement}: \textit{at each fixed value of the parameter $\gamma_1$, there is a critical value of the interlayer interaction parameter $W_c$ above which the hybridization gap opens in the BLG, and when the hybridization gap is present for a certain value of $\gamma_1$ then it is possible to find a realistic critical value of the parameter $\gamma_1$ itself, at which the hybridization gap closes, rendering the BLG into the possible excitonic condensate state.}       
These statements are not valid only in the case of the very large interlayer interactions, for example at $W=2\gamma_0$, at which the very high, but approximatively realistic (for example $\gamma_1=0.5\gamma_0$), values of the interlayer hopping amplitude are not capable of suppressing the very large hybridization gap. Indeed, we think that the charge neutrality at very large $W$ is rather not realistic and could not be achieved experimentally by anyway.   

One of the principal achievements, which also ensues from our theoretical model, is the existence of the excitonic condensate states in the BS type bilayer graphene mediated by the interlayer hopping amplitude, even at the finite and relatively small values of the interlayer interaction parameter. The density of states calculations, effectuated in the present paper, show that the excitonic condensate and the excitonic pair formation are fully controlled by the interlayer Coulomb interaction and interlayer hopping. Moreover, in the limit $W\neq0$, there exists an interesting inter-crossover from the hybridized insulating gapped state to the excitonic condensate states in the BLG, mediated by the interlayer hopping mechanism. Therewith, we have shown that the passage when $W=0$, is not strictly speaking equivalent to the usual tight-binding description of the BLG.
 
The different interlayer interaction regimes have been considered in the paper, which correspond to different screening regimes in the bilayer graphene, and which have been discussed only partially in the known literature. 
From the experimental side of the problem, and taking into account the recent theoretical achievements on the bilayer graphene systems, \cite{cite_4, cite_9,cite_10, cite_11, cite_12}, the Coulomb drag measurements \cite{cite_28, cite_29, cite_30} are promising to observe the excitonic condensation in the pure BLG (without strong disorder) and double BLG heterostructures.
For the future, the study the excitonic effects in the hBN intercalated multilayer graphene $G/hBN/G$ could have a breakthrough impact in the technological applications and improvements of these materials as the solid state systems with the sufficiently large band gaps and also due to the recently growing interests in these materials for the potential interconnected circuit technologies with the improved high current capacities across these structures, approaching the pristine graphene's working performances.
 
%
\section{\label{sec:Section_7} Author contribution statement}
%
All authors contributed equally to the paper. 

\appendix


%
\end{document}